\documentclass[A4,12pt]{article}
\usepackage{jcappub}

\begin{document} 
\title {Observational Constraints on EoS parameters of Emergent Universe }

\author[a]{B. C. Paul,}
\author[b]{P. Thakur}

\affiliation[a]{Physics Department, North Bengal University \\
      Dist. : Darjeeling, Pin : 734 013, West Bengal, India }

\affiliation[b]{Physics Department, Alipurduar College \\ 
      Dist. : Jalpaiguri, Pin : 736122, West Bengal, India }

\emailAdd{bcpaul@iucaa.ernet.in} 
\emailAdd{thakur1 @ yahoo.co.in}

\abstract{
We investigate emergent universe model using recent observational data for the  background tests as well as  for the growth tests. The background test data comprises of Hubble data, Baryon Acoustic Oscillation (BAO) data, cosmic microwave background (CMB) shift data, Union compilation data. The observational growth data are obtained from Wiggle-Z measurements and  rms mass fluctuations data from Lyman-$\alpha$ measurements at different red shifts. The flat emergent universe model obtained by Mukherjee {\it et. al.} is permitted with a  non-linear equation of state (in short, EoS) ($p=A\rho-B\rho^{\frac{1}{2}}$), where $A$ and $B$ are constants. The observed cosmological data  are used here to estimate the range of  allowed values of  EoS parameters numerically. The    best-fit values of the EoS  and growth parameters are determined making use of chi-square minimization technique. The analysis is carried out here considering the Wang-Steinhardt ansatz for  growth index ($\gamma$) and growth function ($f$ defined as $f=\Omega_{m}^{\gamma} (a)$).  Subsequently, the best-fit values of the EoS parameters are used to study the evolution of  the  growth function $f$, growth index $\gamma$, state parameter $\omega$  with red shift parameter $z$.  The present value of the parameters,  namely, $f$, $\gamma$, $\omega$, $\Omega_{m}$ are also estimated. The late accelerating phase of the universe in  the model is accommodated satisfactorily. }

\maketitle
\flushbottom

\section{Introduction}
\label{sec:introd}
The recent cosmological and astronomical observations predict that the present universe is expanding. It is also believed that the present universe might have emerged out of an inflationary phase in the past. After the discovery of Cosmic Microwave Background Radiation  \cite{penin, dicke} the big-bang model become the standard model of the universe which has a beginning of the Universe at some finite past. However, big-bang model based on perfect fluid  fails to account  some of the observed facts of the universe. Further it is observed that while probing the early universe a number of problems namely, the horizon problem, flatness problem, singularity problem, large scale structure formation problem cropped up.  In order to resolve those issues  of the early universe the concept of inflation\cite{guth81,sato81,linde82,stein82}  in cosmology was introduced.
A number of inflationary models are proposed in the last thirty years. The recent cosmological observations predict that the present   universe is passing through a phase of acceleration \cite{sn1,sn4,perl98,perl99} another mystery of the universe. This phase of acceleration is believed to be a late time expansion phase of the universe which can be accommodated in the standard model with the help of a positive cosmological constant.  However, the physics of the inflation and introduction of a small cosmological constant for late time acceleration, is not clearly understood \cite{albrec00,carol01}. In the literature  the late accelerating phase of the universe  is obtained with exotic matter or with a modification of the Einstein gravity. A non-linear equation of state  is also considered in the literature to construct cosmologies \cite{euorg}, emergent universe model is one such model. The Emergent universe (EU) model obtained by Mukherjee {\it et al.} in the flat universe permits an accelerating phase.

Emergent universe scenario was introduced mainly to avoid the initial singularity. It replaces the initial singularity by an Einstein static phase in which the scale factor of the Friedmann-Robertson-Walker (FRW) metric does not vanish. As a result  the energy density, pressure  do not diverge. In this description the universe started expanding from the initial phase, smoothly joins a stage of exponential inflation followed by standard reheating then subsequently it approaches the classical thermal
radiation dominated era of the conventional big bang cosmology \cite{ellis}. Universe in this model can stay large enough to avoid quantum gravitational effects even in the very early universe.

The emergent universe (EU)  scenario is considered to begin from a static Einstein universe forever  which  afterwards successfully  accommodates the early inflationary phase and avoid the messy situation of the initial singularity \cite{ellis,har}. In EU  model  late time de-Sitter phase exists which naturally incorporates the late time accelerating phase as well. EU scenario has been  explored with quintom matter \cite{ca} and further investigated the realization of the scenario with a non-conventional fermion field to obtain a scale invariant perturbation \cite{ca1}. It has been shown that the EU scenario can be implemented successfully  in the framework of general relativity \cite{euorg} in addition to   Gauss-Bonnet gravity \cite{eugb}. The modified Gauss-Bonnet gravity as gravitational alternative for dark energy is however considered by Nojiri and Oddintsov \cite{no}.  It is also shown recently that EU model can be successfully implemented in  Brane world gravity \cite{baner08,deb08}, Brans-Dicke theory \cite{eubd}. A number of cosmological models are obtained with different cosmological fluids and fields \cite{ellis,snp1,snp2,snp3,varun} where initial singularity problem is addressed.   Mukherjee {\it et al.} \cite{euorg} obtained an emergent universe model in the framework of  GTR with a polytropic equation of state (EoS) given  by 
\begin{equation}
\label{euEoS}
p=A\rho-B\rho^{\frac{1}{2}}
\end{equation}
where $A$, $B$ are  constants with $B>0$. It is interesting as it avoids  the initial singularity problem and the initial size of the universe is large. EU model also accommodates the late accelerating phase. 
It may be pointed out here that the EoS state parameters in the model play an important role which decide the composition of matter in the universe. 
In Ref. \cite{euorg}, it is shown that for discrete set of values of A namely, $A(=0,-1/3,1/3,1)$, one obtains universe with a mixture  of three different kinds of cosmic fluids.  The dark energy  is one of the prime constituents of the mixture.  Each of the above EU model have dark energy as one of its constituent fluid. The parameters $A$ and $B$ are arbitrary. Therefore we study here EU model to determine the parameters from cosmological observational aspects. \\ 
The analysis we adopt here consists of both the background test and the growth test.

(A):Using background tests :

There are four main background tests for a cosmological model:
1. The differential age of old galaxies, given by H(z).
2. The  CMB shift parameter.
3. The peak position of the baryonic acoustic oscillations (BAO).
4. The SN Ia data.

We use $H(z)-z$ data \cite{stern10} given in table-\ref{tab1}. The supernovae data is taken from  the union compilation data \cite{kowal08}.

(B): Using  growth data :

 The growth of the large scale structures derived from linear matter density contrast $\delta(z)\equiv\frac{\delta\rho_{m}}{\rho_{m}}$ of the universe is considered to be an  important tool to constrain cosmological model parameters. In this case one parametrizes the growth function $f=\frac{d\log\delta}{d\log a}$ in terms of growth index $\gamma $ to describe the evolution of the inhomogeneous energy density. Initially, Peebles \cite{peebles} and later Wang and Steinhardt \cite{wangstein} parametrized $\delta$ in terms of $\gamma$. The above parametrizations in cosmology have been used in  different contexts in the literature \cite{linder,ja,lue,lue3, dan}.\\ 

 The growth data set given by table(\ref{tab2})  corresponds to a value for growth function $ f$  at a given red-shift. In table(\ref{tab3})  growth data displayed  corresponds to  various sources such as: the red-shift distortion of galaxy power spectra \cite{hawkins1}, root mean square $(rms)$  mass fluctuation ($\sigma_{8}(z)$) obtained from galaxy and Ly-$\alpha$ surveys at various red-shifts \cite{viel1,viel2}, weak lensing statistics \cite{kaiser}, baryon acoustic oscillations \cite{bao05}, X-ray luminous galaxy clusters \cite{manz}, Integrated Sachs-Wolfs (ISW) Effect \cite{rees,am,kaiser3,cr,pog}.\\

It is known that red-shift distortions are caused by velocity flow induced by gravitational potential gradient which evolved both due to the growth of the universe under gravitational attraction  and dilution of the potentials due to the cosmic expansion. The gravitational growth index $\gamma$ is also related to red-shift distortions \cite{linder}. The  cluster abundance evolution,  however, strongly depends on rms mass fluctuations ($\sigma_{8}(z)$) \cite{wangstein}e which will be also considered in the present analysis.\\

The paper is presented as follows : In sec.2, relevant field equations obtained from Einstein field equations are given. In sec.3,  constraint on the EoS parameters obtained from background test  are presented. In sec.4, growth index parametrisation in terms of EoS parameters is studied. In sec.5,  constraint on the EoS parameters obtained from background test and growth test  are determined. In sec.6, a summary of the results analysed are tabulated. Finally, in sec.7, we give a brief discussion.       

\section{Field Equations}

The Einstein field equation is given by
\begin{equation}
\label{ricci}
R_{\mu \nu}-\frac{1}{2} g_{\mu \nu} R = 8 \pi G \; T_{\mu \nu}
\end{equation}
where $R_{\mu \nu}$ represents Ricci tensor, $R$ represents Ricci scalar, $T_{\mu \nu}$ represents energy momentum tensor and $g_{\mu \nu}$ represents the metric tensor in 4-dimensions.
We consider a Robertson-Walker  metric which is given by
\begin{equation}
\label{metric}
ds^{2} = - dt^{2} + a^{2}(t) \left[ \frac{dr^{2}}{1- k r^2} + r^2 ( d\theta^{2} + sin^{2} \theta \;
d  \phi^{2} ) \right]
\end{equation}
where  $k=0,+1(-1)$ is the curvature parameter in the spatial section representing flat, closed (open) universe respectively and $a(t)$ is the scale factor of the universe with $r,\theta,\phi$  the dimensionless co-moving co-ordinates.

Using metric (\ref{metric}) in the Einstein field eq.(\ref{ricci}),  we obtain the following equations:
\begin{equation}
\label{fried}
3 \left( \frac{\dot{a}^2}{a^2} + \frac{k}{a^2} \right) = 8 \pi G \; \rho, 
\end{equation}
\begin{equation}
2 \frac{\ddot{a}}{a} + \frac{\dot{a}^2}{a^2} + \frac{k}{a^2} = - 8 \pi G \; p,
\end{equation}
where $\rho$ and $p$ represent the energy density and pressure respectively. The conservation equation is given by
\begin{equation}
\label{energy}
\frac{d\rho}{dt} + 3 H \left(\rho + p \right) = 0, 
\end{equation}
where $H = \frac{\dot{a}}{a}$ is Hubble parameter.
Using EoS given by eq.(\ref{euEoS}) in eq.(\ref{energy}), and integrating once we obtain  energy density  which is given by 
\begin{equation}
\label{rho}
\rho_{eu}=\left[\frac{B}{1+A}+\frac{1}{A+1}\frac{K}{a^{\frac{3(A+1)}{2}}}\right]^{2}
\end{equation}
where K is a positive integration constant. For convenience we rewrite  eqn(\ref{rho})  as
\begin{equation}
\label{rhoeu}
\rho_{eu}=\rho_{eu0}\left[A_{s}+\frac{1-A_{s}}{a^{\frac{3(A+1)}{2}}}\right]^{2}
\end{equation}
where $ A_{s} = \frac{B}{1+A}\frac{1}{\rho_{eu0}^{\frac{1}{2}}}$ and $\frac{K}{A+1}=\rho_{eu0}^{\frac{1}{2}}-\frac{B}{A+1}$. The scale factor of the universe can be expressed as  $\frac{a}{a_{0}}=\frac{1}{1+z}$, where   $z$ is the red-shift parameter and  we choose the present scale factor of the universe   $a_{0}=1$. 
Therefore the Hubble parameter  in terms of red-shift parameter can be rewritten using the field  eq. (\ref{fried}) as
\begin{eqnarray}
\label{hpara}
H(z)=H_{0}[\Omega_{b0}(1+z)^3+ 
(1-\Omega_{b0})
(A_{s}+(1- A_{s})(1+z)^{\frac{3(A+1)}{2}})^2]^{\frac{1}{2}}
\end{eqnarray}
where $\Omega_{b0}$, $H_{0}$ represents the present baryon density and  Hubble parameter respectively.
The square of the  speed of sound is given by
\begin{equation}
\label{sound}
c^2_{s}=\frac{\delta p}{\delta\rho}=\frac{\dot{p}}{\dot{\rho}}
\end{equation}
which reduces to
\begin{equation}
\label{sound1}
c^2_{s}=A-\frac{A_{s}(1+A)}{2(A_{s}+(1- A_{s})(1+z)^{\frac{3(A+1)}{2}})}.
\end{equation}
In terms of state parameter it reduces to
\begin{equation}
\label{sound2}
c^2_{s}=\frac{\omega+A}{2}.
\end{equation}
From the above equation we obtain the inequality 
\begin{equation}
\label{spd}
A_{s} < 2 \frac{A}{A+1} 
\end{equation}
for a realistic solution which admits stable perturbation \cite{lxu}. Again positivity of sound speed  leads to a upper bound on $c_{s}^2 \leq 1$ which arises from the causality condition.

The deceleration parameter is given by
\begin{equation}
q(a)=\frac{\frac{\Omega_{b0}}{a^3}+\Omega_{eu}(a)[1+3\omega(a)]}{2[\frac{\Omega_{b0}}{a^3}+\Omega_{eu}(a)]}
\end{equation}
where
\begin{equation}
\Omega_{eu}(a)=\Omega_{eu0}[A_{s}+\frac{(1- A_{s})}{a^{\frac{3(A+1)}{2}}}]^2
\end{equation}

\section{Background tests}
 
We consider the following  background tests from observed cosmological data for  analyzing cosmological models: \\
1. The differential age of old galaxies, given by $ H(z)$.
2. The $ CMB $ shift parameter.
3. The peak position of the baryonic acoustic oscillations $(BAO)$.
4. The $SN Ia$ data.

\subsection{Observational Constraints}
The equation of state for emergent universe contains two unknown parameters namely $A$ and $B$, which are  determined from numerical analysis adopted here for different observed data. For this the Einstein field equation is  rewritten in terms of a dimensionless Hubble parameter and a suitable chi-square function is defined in different cases. 

Case I: For $OHD$ \\
The observed Hubble Data is then taken from the table given below \cite{stern10} :
To analyze first we define  chi-square $\chi^2_{H-z}$ function  is given by
\begin{equation}
\chi^{2}_{H-z}(H_{0},A_{s},A,z)=\sum\frac{(H(H_{0},A_{s},A,z)-H_{obs}(z))^2}{\sigma^{2}_{z}}
\end{equation} 
where $H_{obs}(z)$ is the observed Hubble parameter at red shift $z$ and $\sigma_{z}$ is the error associated with that particular observation  as shown  in table -\ref{tab1}. 
\begin{table}
  \centering
  \begin{tabular}{|l|r|c|c|}
  \hline
  {\it z } & $H(z)$ & $\sigma$ \\
  \hline
   0.00 & 73  & $ \pm $ 8.0	 \\
   0.10 & 69  & $ \pm $ 12.0 \\
   0.17 & 83  & $ \pm $ 8.0 \\
   0.27 & 77  & $ \pm $ 14.0 \\
   0.40 & 95  & $ \pm $ 17.4 \\
   0.48 & 90  & $ \pm $ 60.0 \\
   0.88 & 97  & $ \pm $ 40.4 \\
   0.90 & 117 & $ \pm $ 23.0 \\
   1.30 & 168 & $ \pm $ 17.4 \\
   1.43 & 177 & $ \pm $ 18.2 \\
   1.53 & 140 & $ \pm $ 14.0 \\
   1.75 & 202 & $ \pm $ 40.4 \\

\hline
\end{tabular}
\caption{\label{tab1} $H(z) vs. z$ data from Stern {\it et al. } \cite{stern10}}
\end{table} 

Case II : For $BAO $ //

 A model independent $BAO$ (Baryon Acoustic Oscillation) peak parameter  for low red shift $z_{1}$ measurements in a flat universe is given by \cite{bao05}:
\begin{equation}
\label{baop20}
\mathcal {A} =\frac{\sqrt{\Omega_{m}}}{E(z_{1})^{1/3}}\left(\frac{\int ^{z_{1}}_0 \frac{dz}{E(z)}}{z_{1}}\right)^{2/3} 
\end{equation}
where $\Omega_{m}$ is the matter density parameter for the Universe. The chi square function in this case is  defined as :
\begin{equation}
\label{chibao20}
\chi^{2}_{BAO}(A_{s},A,z)=\frac{\left(\mathcal{A}-0.469\right)^{2}}{ \left(0.017\right)^{2}}.
\end{equation}
The SDSS data for Luminous Red Galaxies (LRG) survey gives  $\mathcal{A}$ ($0.469\pm.0.017$)  \cite{bao05}.

Case : III $CMB $ //

The $CMB$ shift parameter ($\mathcal {R}$) is given by \cite{komatsu10}:
\begin{equation}
\label{cmbp20}
\mathcal{R}=\sqrt{\Omega_{m}}\int ^{z_{ls}}_{0} \frac{dz'}{H(z')/H_{0}}
\end{equation}
where $z_{ls}$ is the $z$ at the surface of last scattering. The WMAP7 data predicts $\mathcal{R}=1.726 \pm 0.018$ at $z=1091.3$. We now define chi-square function  as :
\begin{equation}
\label{chicmb20}
\chi^2_{CMB}(A_{s},A,z)=\frac{(\mathcal{R}-1.726)^2}{(0.018)^2}.
\end{equation}

\begin{figure}
\centering
{\includegraphics[width=8cm,height=6cm]{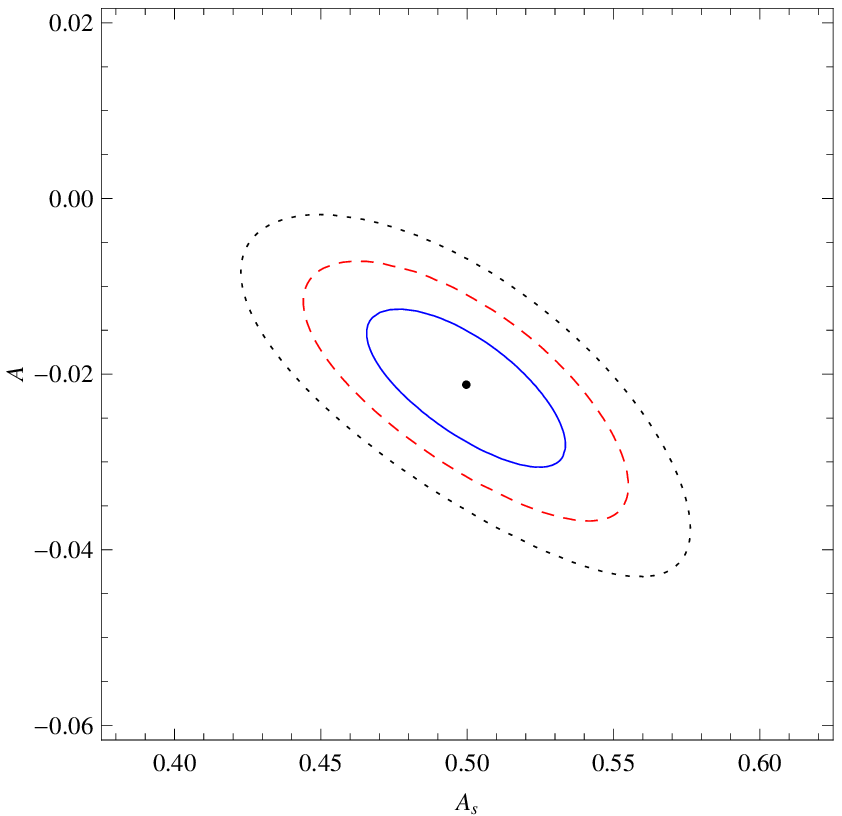}}\\
{\includegraphics[width=8cm,height=5cm]{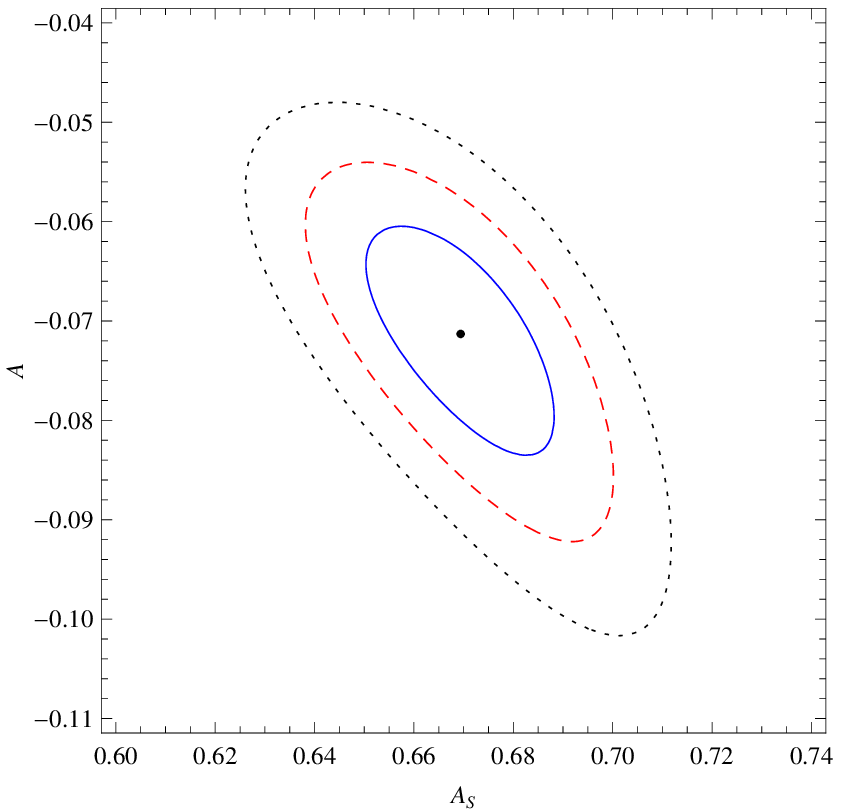}}\\
{\includegraphics[width=8cm,height=5cm]{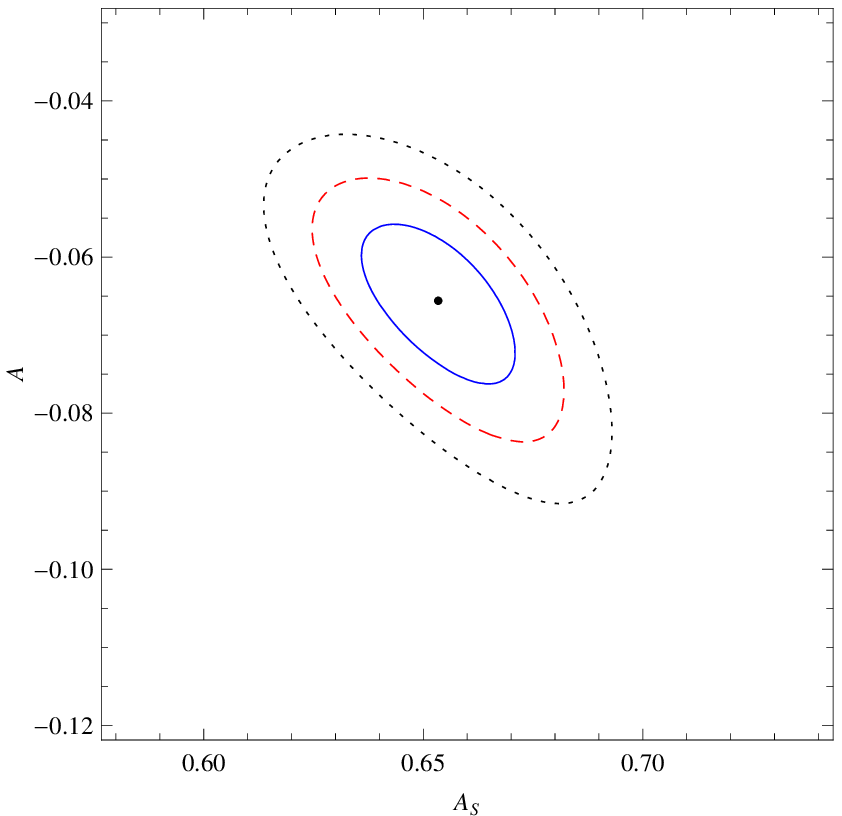}}\\
\caption{Contours of $A_{s}$ and $A$ from(i) H-z+BAO+CMB shift data(ii) at  H-z+BAO+CMB+Union2 (iii) H-z+BAO+CMB+Union2+growth+rms mass fluctuation data  at WMAP7 Hubble value at  $68.3$\%(Solid) $95.4\%$ (Dashed) and $99.7 \%$  (Dotted) confidence limit }
\label{wmap7}
\end{figure}

\begin{figure}
\centering
{\includegraphics[width=8cm,height=6cm]{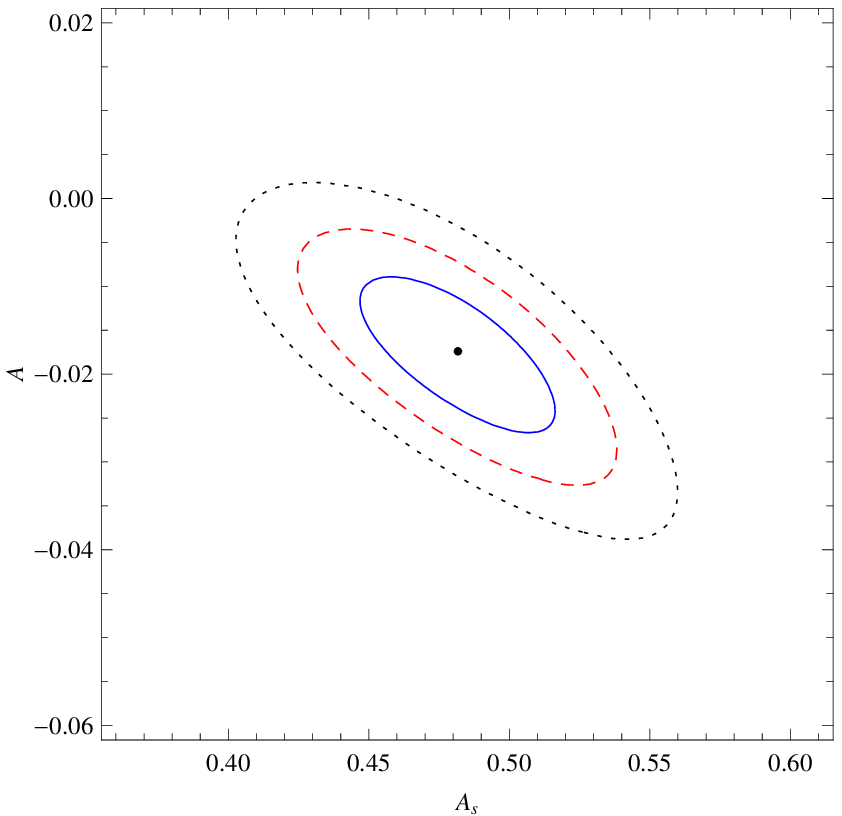}}\\
{\includegraphics[width=8cm,height=5cm]{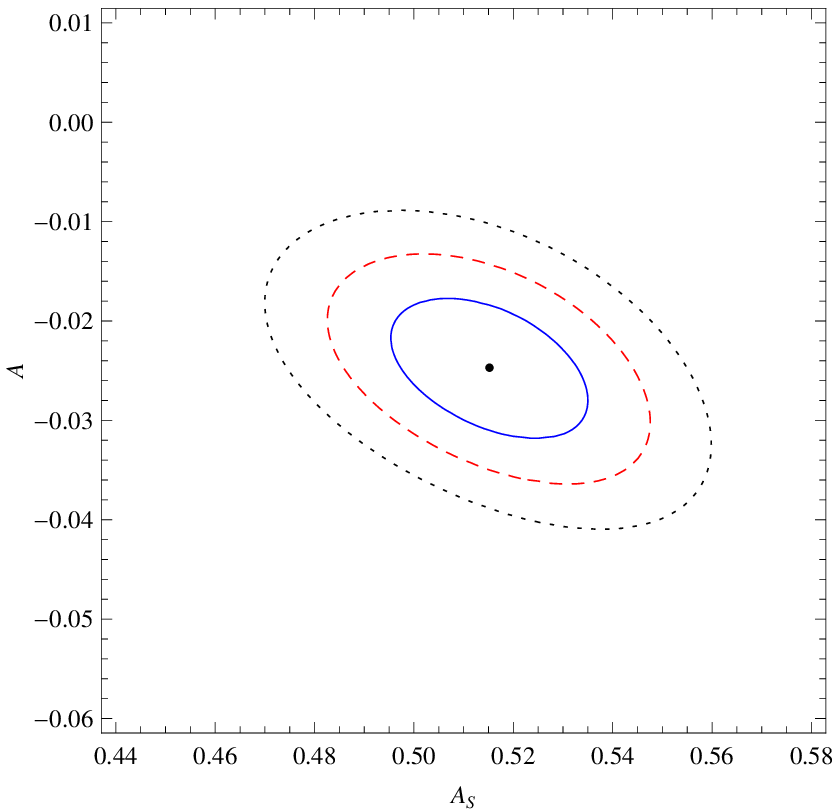}}\\
{\includegraphics[width=8cm,height=5cm]{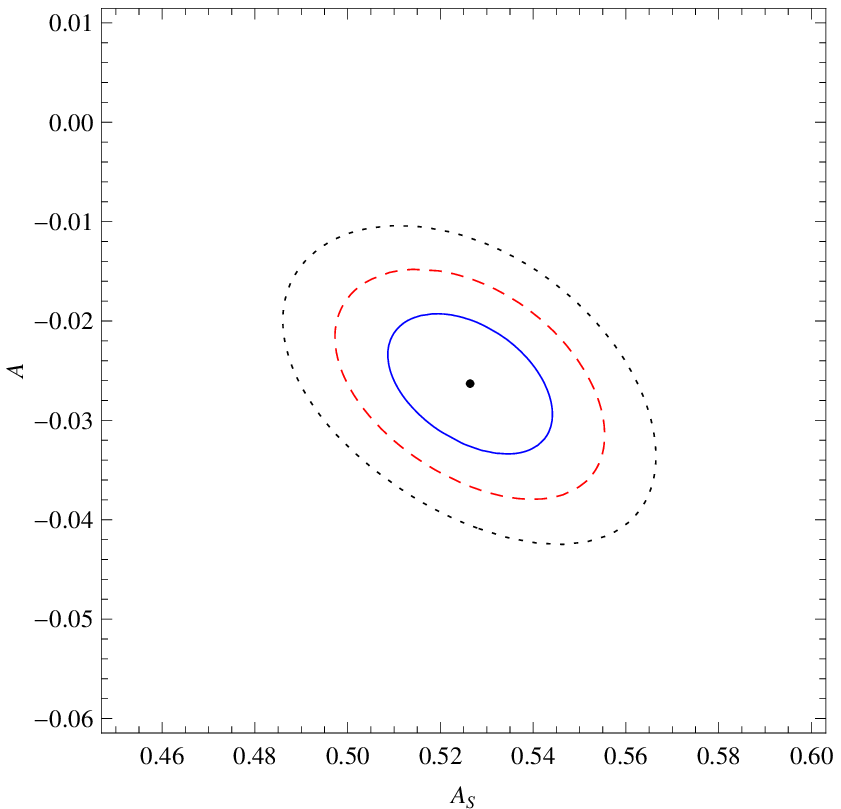}}\\
\caption{Contours of $A_{s}$ and $A$ from(i) H-z+BAO+CMB shift data(ii) at  H-z+BAO+CMB+Union2 (iii) H-z+BAO+CMB+Union2+growth+rms mass fluctuation data  at Planck 2013 Hubble value at  $68.3$\%(Solid) $95.4\%$ (Dashed) and $99.7 \%$  (Dotted) confidence limit }
\label{planck13}
\end{figure}

Now we combine the above three chi-square functions to define a total chi-square function  as  $\chi ^2_{hbc} = \chi ^2_{H-z}+\chi ^2_{BAO}+\chi ^2_{CMB}$. The above function is minimized at the present Hubble value as predicted by  WMAP7 and PLANCK2013 \cite{planck13}. The best fit values of $A$ and $A_s$ are also determined.  The  contours between $A$ and $A_{s}$ are drawn at different confidence level  which are shown in fig (\ref{wmap7}a,\ref{planck13}a).

Case IV : $Supernova$

The distance modulus function ($\mu$) is defined  in terms of luminosity distance ($d_{L}$)  as
\begin{equation}
\label{mu10}
\mu(A_{s},A,z)= m-M = 5\log_{10}(d_{L})
\end{equation}
where 
\begin{equation}
\label{lum10}
d_{L}=\frac{c(1+z)}{H_{0}} \int ^{z}_{0} \frac{dz'}{E(z')}
\end{equation}

In this case the chi-square $\chi^2_{\mu}$ function  is defined as
\begin{equation}
\chi^{2}_{\mu}(A_{s},A,z)=\sum\frac{(\mu(A_{s},A,z)-\mu_{obs}(z))^2}{\sigma^{2}_{z}}
\end{equation} 
where $\mu_{obs}(z)$ is the observed distance modulus at red shift $z$ and $\sigma_z$ is the corresponding error for the observed data\cite{kowal08}.
Finally the chi-square function for background tests is defined as
\begin{equation}
\label{chiback}
\chi^{2}_{back}(A_{s},A,z)=\chi^2_{H-z}(A_{s},A,z) +\chi ^2_{BAO}(A_{s},A,z)+\chi ^2_{CMB}(A_{s},A,z)+\chi^{2}_{\mu}(A_{s},A,z)
\end{equation}

The  chi-square function for background test is minimized with present Hubble value predicted by WMAP7, PLANCK2013. The best fit values of $A$ and $A_{s}$ are determined.  The contours  between $A$  and $A_{s}$ are drawn at different confidence level  which are shown in figs.(\ref{wmap7}b,\ref{planck13}b).

\section{Parametrization of the Growth Index}

In this section the growth rate of the large scale structures is derived from matter density perturbation given  by $\delta=\frac{\delta\rho_{m}}{\rho_{m}}$ (where  $\delta\rho_{m}$ represents the fluctuation of matter density $\rho_{m}$) in the linear regime which satisfies the following equation \cite{padma,liddle99}:
\begin{equation}
\label{grrate}
\ddot{\delta}+2\frac{\dot{a}}{a}\dot{\delta}-4\pi G\rho_{m}\delta=0.
\end{equation}
The field equations for the background cosmology in a flat Robertson-Walker metric are given below
\begin{equation}
\left(\frac{\dot{a}}{a}\right)^{2}=\frac{8\pi G}{3}(\rho_{b}+\rho_{eu}),
\end{equation}
\begin{equation}
2\frac{\ddot{a}}{a}+\left(\frac{\dot{a}}{a}\right)^{2}=- 8\pi G \omega_{eu}\rho_{eu}
\end{equation}
where  $\rho_{eu}$  and  $\omega_{eu}$   represent  energy density and the equation of  state parameter for  EU respectively  with  $\rho_{b}$ as the  background energy density. The equation of  state parameter for EU corresponding to the EoS given by eq. ( \ref{euEoS} ) is
\begin{equation}
\omega_{eu}=A-\frac{A_{s}(1+A)}{(A_{s}+(1- A_{s})(1+z)^{\frac{3(A+1)}{2}})}.
\end{equation}
Now we replace time variable ( $t$ ) by a scale factor variable  in the aboveto solve the equation. Consequently, we replace  $t$ variable to $\ln a$ in  eq.(\ref{grrate}),  and finally  obtain
\begin{equation}
\label{delta}
(\ln\delta)^{''}+(\ln\delta)^{'2}+(\ln\delta)^{'}
 \left[\frac{1}{2}-\frac{3}{2}\omega_{eu}(1-\Omega_{m}(a))\right]=\frac{3}{2}\Omega_{m}(a)
\end{equation}
where
\begin{equation}
\label{density}
\Omega_{m}(a)=\frac{\rho_{m}}{\rho_{m}+\rho_{eu}}.
\end{equation}
The  effective matter density is given by 
\begin{equation}
\label{density1}
\Omega_{m}(a)=\frac{H^2_{0}\Omega_{0m}a^{-3}}{H^2(a)}
\end{equation}
where $\Omega_{0m}=\Omega_{b}+(1-\Omega_{b})(1-A_{s})^2$ \cite{li}.
Using the energy conservation eq. (\ref{energy}) and  changing   $\ln a$ variable to $\Omega_{m}(a)$ once again it is possible to rewrite   the eq. (\ref{delta})  in terms of the logarithmic growth factor ($f=\frac{d\log\delta}{d\log a}$), which is given by 
\begin{equation}
\label{gfactor}
3\omega_{eu}\Omega_{m}(1-\Omega_{m})\frac{d f}{d \Omega_{m}}+f^{2}
+f\left[\frac{1}{2}-\frac{3}{2}\omega_{eu}(1-\Omega_{m}(a))\right]=\frac{3}{2}\Omega_{m}(a).
\end{equation}
The logarithmic growth factor $f$, according to Wang and Steinhardt   \cite {wangstein} is given by
\begin{equation}
\label{ansatzf}
f=\Omega_{m}^{\gamma}(a)
\end{equation} 
where $\gamma$ represents the growth index parameter. In the case of flat dark energy model  with constant state parameter  $\omega_{0}$, the growth index $\gamma$ is given by
\begin{equation}
\label{gamflat}
\gamma=\frac{3(\omega_{0}-1)}{6\omega_{0}-5}.
\end{equation}
For a  $\Lambda CDM$ model, it reduces to $\frac{6}{11}$ \cite{linder,evl}, however, for a matter dominated model one obtains $\gamma=\frac{4}{7}$ \cite{fry,ne}.  It is also convenient to express  $\gamma$  as a parametrized function of red shift parameter $z$ in cosmology.   One  of the parametrized form  is obtained from the Taylor expansion of the function about $z=0$ keeping the first two terms only. Accordingly one obtains
\[
\gamma(z)=\gamma(0)+\gamma^{'}z, 
\]
where $\gamma^{'}\equiv\frac{d\gamma}{dz}|_{(z=0)}$ \cite{polarski, gan}. It has been shown recently \cite{ishak} that it smoothly interpolates a low and intermediate red shift range to a high red shift range up to the cosmic microwave background (CMB)  scale. Similar parametrization technique is also used in cosmology in different contexts \cite{dosset} to study evolution.
Here we parametrize  $\gamma$  in terms of EoS parameters for emergent universe namely,  $A$ and $A_{s}$.
 Therefore, we begin with the following  ansatz which is given by
\begin{equation}
\label{ansatzeu}
f=\Omega_{m}^{\gamma(\Omega_{m})} (a)
\end{equation} 
where the growth index parameter is represented by $\gamma(\Omega_{m})$. It can be expanded in Taylor series expansion around $\Omega_{m} = 1$ which leads to 
\begin{equation}
\label{tylor}
\gamma(\Omega_{m})=\gamma|_{(\Omega_{m}=1)}
 +(\Omega_{m}-1)\frac{d\gamma}{d\Omega_{m}}|_{(\Omega_{m}=1)} + O(\Omega_{m}-1)^{2}.
\end{equation}
Now the eq.  (\ref{gfactor}) can be rewritten in terms of $\gamma$ as
\begin{equation}
\label{tylomega}
3\omega_{eu}\Omega_{m}(1-\Omega_{m})\ln\Omega_{m}\frac{d\gamma}{d \Omega_{m}}-3\omega_{eu}\Omega_{m}(\gamma-\frac{1}{2})+ \Omega_{m}^{\gamma}-\frac{3}{2}\Omega_{m}^{1-\gamma}+3\omega_{eu}\gamma-\frac{3}{2}\omega_{eu}+\frac{1}{2}=0.
\end{equation}
Differentiating once again the above equation around $\Omega_{m}=1$, one obtains a zeroth order term in the expansion  for $\gamma$  given by
\begin{equation}
\label{greu}
\gamma=\frac{3(1-\omega_{eu})}{5-6\omega_{eu}},
\end{equation}
which  supports a dark energy model for a constant $\omega_{0}$ (eq. \ref{gamflat}).

Differentiating the above equation once again with respect to $\Omega_{m}$, the first order terms  in the expansion at $\Omega_{m}=1$, is given by
\begin{equation}
\label{forder}
\frac{d\gamma}{d\Omega_{m}}|_{(\Omega_{m}=1)} = \frac{3(1-\omega_{eu})(1-\frac{3\omega_{eu}}{2})}{125(1-\frac{6\omega_{eu}}{5})^{3}}.
\end{equation}
Using the above equation in  eq. (\ref{tylor}), $\gamma$ is further determined. Now the zeroth and first order terms together give the following expression
\begin{equation}
\label{gammaeu}
\gamma=\frac{3(1-\omega_{eu})}{5-6\omega_{eu}}+(1-\Omega_{m})\frac{3(1-\omega_{eu})(1-\frac{3\omega_{eu}}{2})}{125(1-\frac{6\omega_{eu}}{5})^{3}}.
\end{equation}
Using the expression of $\omega_{eu}$ in the above,  $\gamma$ can be parametrised in EU model in terms of the EoS parameters, namely,  $A_{s}$, $A$ respectively  and red shift parameter $z$ .

 Let us now  define normalized growth function $g$ by  
 \begin{equation}
 \label{g}
 g(z)\equiv\frac{\delta(z)}{\delta(0)}
 \end{equation}
 which is determined from the solution of  eq. (\ref{delta}).
 Thereafter the corresponding  approximate normalised growth function is obtained from the parametrized form of $f$  from eq.(\ref{ansatzeu}). It  is given by
 \begin{equation}
 g_{th}(z)=\exp\left[ \int_{1}^{\frac{1}{1+z}}  \Omega_{m}(a)^{\gamma}\frac{da}{a} \right]
 \end{equation}
  which will be considered here to construct  chi-square function in the next section.

\subsection{Observational Constraints}

\begin{table}
\centering
		\begin{tabular}{@{}|l|c|c|r|}
		\hline
		z     & $f_{obs}$ &    $\sigma$ &    $Ref.$\\
		\hline
		$0.15$   & 0.51 &  0.11 &  \cite{hawkins,verde}\\
		$0.22$ & 0.60 & 0.10 &  \cite{blake}\\
		$0.32$ & 0.654 & 0.18 & \cite{reyes}\\
		$0.35$ & 0.70 & 0.18 &  \cite{tegmark}\\
		$0.41$ & 0.70 & 0.07 & \cite{blake}\\
		$0.55$ & 0.75 & 0.18 & \cite{ross}\\
		$0.60$ & 0.73 & 0.07 & \cite{blake}\\
		$0.77$ & 0.91 & 0.36 & \cite{guzzo}\\
		$0.78$ & 0.70 & 0.08 & \cite{blake}\\
		$1.4$ & 0.90 & 0.24 & \cite{angela}\\
		$3.0$ & 1.46 & 0.29 & \cite{mcdon}\\
		\hline	
		\end{tabular}
\caption{Data for the observed growth functions $f_{obs}$ used in our analysis }	
	\label{tab2}
\end{table}

\begin{table}
\centering
		\begin{tabular}{@{}|l|c|c|r|}
		\hline
		z     & $\sigma_{8}$ &    $\sigma_{\sigma_{8}}$ &    $Ref$\\
		\hline
		$2.125$   & 0.95 &  0.17 &  \cite{viel1}\\
		$2.72$ & 0.92& 0.17 &       \\
		$2.2$ & 0.92 & 0.16 & \cite{viel2}\\
		$2.4$ & 0.89 & 0.11 &   \\
		$2.6$ & 0.98 & 0.13 &   \\
		$2.8$ & 1.02 & 0.09 & \\
		$3.0$ & 0.94 & 0.08 & \\
		$3.2$ & 0.88 & 0.09 & \\
		$3.4$ & 0.87 & 0.12 & \\
		$3.6$ & 0.95 & 0.16 & \\
		$3.8$ & 0.90 & 0.17 & \\
		$0.35$ & 0.55 & 0.10 & \cite{marin}\\
		$0.6$ & 0.62 & 0.12 &    \\
		$0.8$ & 0.71 & 0.11 &    \\
		$1.0$ & 0.69 & 0.14 &    \\
		$1.2$ & 0.75 & 0.14 &    \\
		$1.65$ & 0.92 & 0.20 &   \\
		\hline	
		\end{tabular}
\caption{Data for the rms mass fluctuations ($\sigma_{8}$) at various red-shift }	
	\label{tab3}
\end{table}

We define chi-square of the growth function $f$ as
\begin{equation}
\chi^{2}_{f}(A_{s},A,z)=\Sigma\left[\frac{f_{obs}(z_{i})-f_{th}(z_{i},\gamma)}{\sigma_{f_{obs}}}\right]^{2}
\end{equation}
where $f_{obs}$ and $\sigma_{f_{obs}}$ are obtained from table (\ref{tab2}). However,  $f_{th}(z_{i},\gamma)$ is obtained from eqs. (\ref{ansatzeu}) and (\ref{gammaeu}).
 Another  observational probe for the matter density perturbation $\delta(z)$ is derived from the red shift dependence of the rms mass fluctuation $\sigma_{8}(z)$.  The dispersion of the density field $\sigma^{2}(R,z)$ on a comoving  scale $R$ is defined as
\begin{equation}
\sigma^{2}(R,z)=\int_{0}^{\inf} W^{2}(kR)\Delta^{2}(k,z)dk/k
\end{equation}
where 
\begin{equation}
W(kR)=3\left(\frac{\sin(kR)}{(kR)^{3}}-\frac{\cos(kR)}{(kR)^{2}}\right)r,
\end{equation}
represents window function , and
\begin{equation}
\Delta^{2}(kz)=4\pi k^{3}P_\delta(k,z),
\end{equation},
where $P_\delta(k,z)\equiv(\delta^{2}_{k})$ is the mass power spectrum at red-shift $z$. The rms mass fluctuation $\sigma_{8}(z)$ is the  $\sigma^{2}(R,z)$  at  $R=8h^{-1}$ Mpc. The function $\sigma_{8}(z)$ is  connected to $\delta(z)$ as
\begin{equation}
\sigma_{8}(z)=\frac{\delta(z)}{\delta(0)}\sigma_{8}|_{(z=0)}
\end{equation}
which implies
\begin{equation}
s_{th}(z_{1}, z_{2}) \equiv \frac{\sigma_{8}(z_{1})}{\sigma_{8}(z_{2})}=\frac{\delta(z_{1})}{\delta(z_{2})}=
\frac{\exp\left[ \int_{1}^{\frac{1}{1+z_{1}}}  \Omega_{m}(a)^{\gamma}\frac{da}{a} \right] }{\exp \left[ \int_{1}^{\frac{1}{1+z_{2}}}  \Omega_{m}(a)^{\gamma}
\frac{da}{a} \right]}.
\end{equation}
In tab-\ref{tab3}, a systematic evolution of rms mass fluctuation $\sigma_{8}(z_{i})$ with observed red shift for flux power spectrum of Ly-$\alpha$ forest \cite{viel1,viel2,marin} are displayed.  In this context we  define  a new chi-square  function which is given by 
\begin{equation}
\chi^{2}_{s}(A_{s},A,z)=\Sigma\left[\frac{s_{obs}(z_{i},z_{i+1})-s_{th}(z_{i},z_{i+1})}{\sigma_{s_{obs,i}}}\right]^{2}.
\end{equation}
Data for rms mass fluctuation at various red shift given in table-\ref{tab3} will be considered here.
Now considering growth function mentioned above, one can  define chi-square function which  is given by 
\begin{equation}
\label{chigr}
\chi^{2}_{growth}(A_{s},A,z)=\chi^{2}_{f}(A_{s},A,z)+\chi^{2}_{s}(A_{s},A,z).
\end{equation}
The chi-square functions defined above will be considered for the analysis in the next section.

\section{Observational constraints from background test and growth test}

Using eq.(\ref{chiback}) and eq.(\ref{chigr}),  we define total chi-square function as
\begin{equation}
\chi^{2}_{total}(A_{s},A,z)=\chi^{2}_{back}(A_{s},A,z)+\chi^{2}_{growth}(A_{s},A,z)
\end{equation}
where
$\chi^{2}_{growth}(A_{s},A,z)=\chi^{2}_{f}(A_{s},A,z)+\chi^{2}_{s}(A_{s},A,z)$.
In this case the best fit values are obtained  minimizing  the  chi-square function. Since chi-square function depends on $A$, $A_{s}$ and $z$, it is possible to draw contours at different confidence limit. The limits imposed by the contours  determines the permitted range of values of the EoS parameters in EU model. The contours  between $A$  and $A_{s}$  are shown in figs.(\ref{wmap7}c,\ref{planck13}c).

\begin{table}[tbp]
\centering
		\begin{tabular}{|l|r|c|c|c|}
		\hline
		$Data$     & $A$ &    $A_{s}$ & $B$ in unit of $\rho_{eu0}$ &  $\chi^{2}/d.o.f$ \\ 	\hline
		$OHD+BAO+CMB$   & -0.0218  & 0.4997 & 0.4888& 0.8974 \\ \hline
		$OHD+BAO+CMB+Union2 $ & -0.0713  & 0.6693 & 0.6216 &1.2137\\ \hline
		$OHD+BAO+CMB+Union2$    &        &       &         &    \\
		$+Growth+\sigma_{8}$ & - 0.0653 &0.6535&0.6108 &1.2039 \\
	  \hline	
		\end{tabular}
\caption{\label{tab4} Best-fit values of the EoS parameters using present Hubble value(WMAP7) }		
\end{table}
\begin{table}
\centering
		\begin{tabular}{|l|r|c|c|c|}
		\hline
		Data &  $CL$  & $A$ & $A_{s}$& $B$ in unit of $\rho_{eu0}$  \\
		\hline
		$OHD+BAO+CMB$ & $99.7\%$  &  $(-0.0433, \;-0.0015)$ & $(0.4221,\; 0.5771)$ &$(0.4038,\;0.5762)$ \\ \hline
		$OHD+BAO+CMB$   &      &                    &                        &                    \\
		$+Union2$ & $99.7\%$ & $(-0.102,-0.0479)$ & $(0.626,0.712)$&$(0.5622,0.6779)$ \\ \hline
		$OHD+BAO+CMB+$   &     &                  &                   &                      \\
		$Union2+Growth+\sigma_{8}$ & $99.7\%$&$(-0.092,\;-0.044)$ & $(0.613,\;0.693)$&$(0.5566,\;0.6625)$\\
	  \hline	
		\end{tabular}
\caption{\label{tab5} Range of  values of the EoS parameters using present Hubble value(WMAP7) }		
	\end{table}

\section{Results}

We analyze the emergent universe model with observed cosmological data. The EoS parameters of emergent universe model is determined numerically using chi-square minimization technique. In this analysis we determine first the best-fit values of the EoS parameters  which are then used to construct a chi-square function. Thereafter the chi-square $\chi^{2}_{total}(A_{s}, A)$ functions are minimized which are then used to draw corresponding contours  connecting $A_{s}$, $A$ at different confidence level. The results are tabulated below :
 
\begin{table}[tbp]
\centering
		\begin{tabular}{|l|r|c|c|c|}
		\hline
		Data     & $A$ &    $A_{s}$& $B$ in unit of $\rho_{eu0}$  &$\chi^{2}/d.o.f$\\
		\hline
		$OHD+BAO+CMB$   & -0.0174 & 0.4817  & 0.4733&0.7779 \\ \hline
		$OHD+BAO+CMB+Union2$ &-0.0247  & 0.515&0.5023 &1.118 \\  \hline
		$OHD+BAO+CMB+Union2$  &         &      &       &\\
		$+Growth+\sigma_{8}$ &- 0.0261  &0.5265 &0.5128& 1.1034 \\
	  \hline	
		\end{tabular}
\caption{\label{tab6} Best-fit values of the EoS parameters using present Hubble value(Planck 2013) }		
\end{table}
\begin{table}
\centering
		\begin{tabular}{|l|r|c|c|c|}
		\hline
		Data &  $CL$  & $A$ & $A_{s}$& $B$ in unit of $\rho_{eu0}$ \\
		\hline
		$OHD+BAO+CMB$ & $99.7\%$   &  $(-0.039,\;0.0020)$ & $(0.403,\; 0.5604)$ &$(0.3873,\;0.5615)$ \\ \hline
		$OHD+BAO+CMB$    &           &                     &                    &      \\
		$+Union2$ & $99.7\%$ & $(-0.0412,\; -0.0087)$ & $(0.47,\; 0.559)$ &$(0.4506,\;0.5541)$ \\ \hline
		$OHD+BAO+CMB+$   &         &                   &                    &                 \\
		$Union2+Growth+\sigma_{8}$ &$99.7\%$&$(-0.0424,\;-0.0101)$ &$(0.486,\;0.557)$&$(0.4654,\;0.5514)$\\
	  \hline	
		\end{tabular}
\caption{\label{tab7} Range of  values of the EoS parameters using present Hubble value(Planck 2013)}		
	\end{table}

\begin{table}
\centering
		\begin{tabular}{|l|r|c|c|c|c|c|c|}
		\hline
		Model &  $A$ & $A_{s}$ &$B$in unit of $\rho_{eu0}$ & $f$ & $\gamma$&$\Omega_{m0}$& $\omega_{0}$\\
		\hline
		
		$EU(WMAP7)$ & $-0.0653$   &  $0.6535$ & $0.6108$ & $0.352$& $0.567$& $0.155$& $-0.678$ \\ \hline
		
		$EU(Planck2013)$ & $-0.0261$ & $0.5265$ & $0.5128$ & $0.454$& $0.572$& $0.255$& $-0.541$ \\   
	  \hline	
		\end{tabular}
\caption{\label{tab10} Values of the EoS parameters in different model}	
	
\end{table}

\begin{figure}
\centering
{\includegraphics[width=230pt,height=200pt]{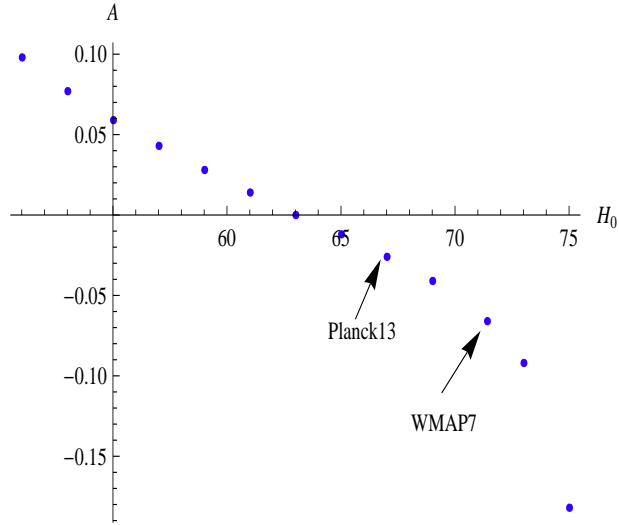}}\\
\caption{Variation of Best-fit values of $A$ with present Hubble value $H_{0}$  }
\label{hub}
\end{figure}

\section{Discussion}

In this paper we present an analysis of   flat emergent universe model \cite{euorg} with observational data.
The analysis of the EU scenario is carried out here numerically by considering (i) background test  and (ii)  combined tests (background + growth test).  Using present observed Hubble value from both WMAP7, Planck 2013 we determine the best-fit values of the parameters  $A$, $B$ in unit of $\rho_{eu0}$  (obtained from  $A_{s}$)  by   chi-square minimization technique. The best fit values of $A$ and $B$ are tabulated in  table-\ref{tab4},\ref{tab6} for WMAP7 and PLANCK2013 respectively. 
Using the best fit values of $A$ and $A_{s}$  contours are  drawn which are shown in figs.(\ref{wmap7}) and (\ref{planck13}). The  permitted range of values of the parameters  for WMAP7 and Planck2013 are then tabulated in table-\ref{tab5},\ref{tab7} respectively. 
  The best-fit values are then employed in EoS, deceleration parameter, growth parameter, growth index  to study the viability of the model. 

\begin{figure}
\centering
{\includegraphics[width=230pt,height=200pt]{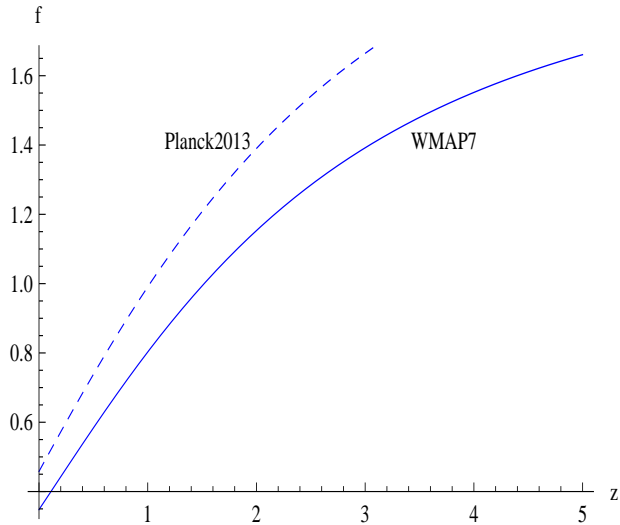}}\\
\caption{Evolution of growth function f with red shift(i) at WMAP7 (solid line),(ii) at Planck 2013 (dashed line) value of present Hubble parameter}
\label{grf}
\end{figure}

\begin{figure}
\centering
{\includegraphics[width=8cm,height=6cm]{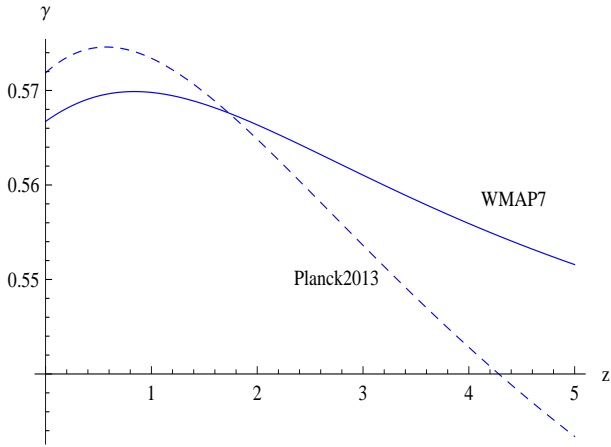}}\\
\caption{Evolution of growth index $\gamma$ with red shift(i) at WMAP7 (solid line),(ii) at Planck 2013 (dashed line) value of present Hubble parameter  }
\label{grind}
\end{figure}

\begin{figure}
\centering
{\includegraphics[width=8cm,height=6cm]{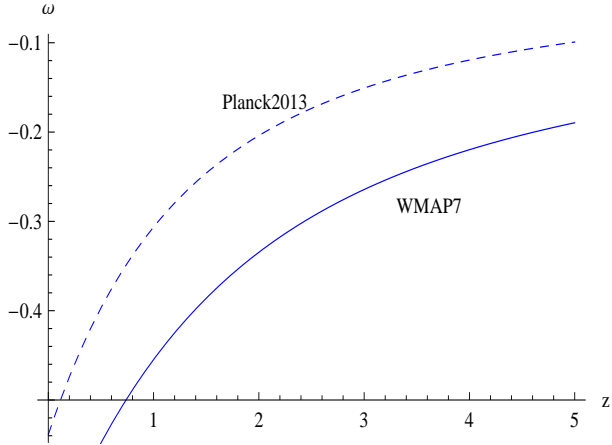}}\\
\caption{Evolution of the state parameter ($\omega$) (i) at WMAP7 (solid line),(ii) at Planck 2013 (dashed line) value of present Hubble parameter  }
\label{omega}
\end{figure}

\begin{figure}
\centering
{\includegraphics[width=8cm,height=6cm]{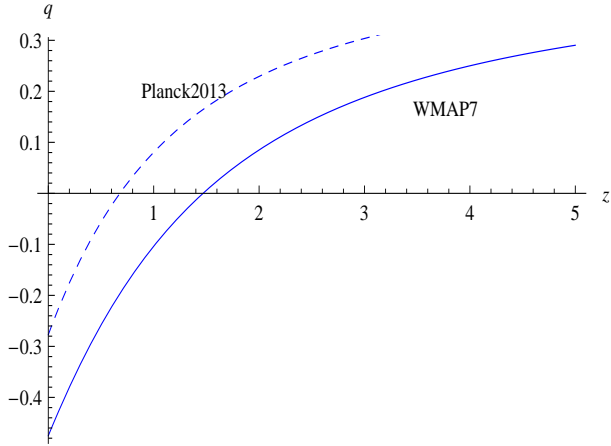}}\\
\caption{Deceleration parameter variations with red shift (i) at WMAP7 (solid line),(ii) at Planck 2013 (dashed line) value of present Hubble parameter  }
\label{decln}
\end{figure}

  We note the following :

As the present Hubble value predicted by PLanck2013 \cite{planck13} is less than that of WMAP7 we consider both the values to analyze the EU model. The best-fit values of the model are $A$=-0.0653, $B$=0.6108 ($A_{s}$=0.6535) for  WMAP7 Hubble value and  $A$=-0.0261, $B$=0.5128 ($A_{s}$=0.5265) for Planck2013  Hubble value.  The lower value for Hubble parameter leads to  an increase in best-fit value for $A$.
Thus we plot a variation of $A$ with different Hubble value in fig. (\ref{hub}). It is found that the change in values of $A$ with $H$ is non-linear but $A$ is found to flip its sign from negative to positive value at a crossover value of $H \sim 63$.
 The present observed data from WMAP7 and Planck2013 both permit an EU scenario with $A \rightarrow 0$ which is evident from the  plot of  state parameter $\omega$, deceleration parameter ($q$) shown in figs. (\ref{omega}) and (\ref{decln}) respectively.\\
 Fig .\ref{grind} is a plot of $\gamma$ with z. It is evident from the value of $\gamma$ that it was small in the early universe and then attained a maximum in the recent past and thereafter decreases slowly. The maximum of $\gamma$ in the case of Planck2013 is more than WMAP7. Also we note that for Plank2013 it attained at lower z value compared to WMAP7.
 Fig.\ref{omega} is the plot of state parameter with z. It shows that $\omega\leq -\frac{1}{3}$ at present  which is also evident from Fig. \ref{decln} that shows the variation of  deceleration parameter with red shift. It is evident  that in the recent past the universe transits from deceleration phase to accelerating phase. Here we see  that there is a flip of sign of q  for z less than one corresponding to planck2013. The analysis carried out here permits a viable cosmological model with the recent Planck2013 data.
Negative values of state parameter (fig.\ref{omega}) in the limit ($\omega \leq$ -1/3) signifies   accelerating  phase of the universe. 
The analysis permits an Emergent universe model with  $A\approx 0$, accommodating dust, dark energy as its constituents. We note that an emergent universe model is more viable in accordance with Planck2013.  
  \\

\section{Acknowledgements}
The authors would like to thank {\it IUCAA Reference Centre} at North Bengal University for extending necessary research facilities to initiate the work. BCP would like to acknowledge  the University Grants Commission (UGC), New Delhi for a Major Research Project Grant (No. F.42-783/(SR) 2013). BCP would like to thank TWAS-UNESCO for awarding Associatship to visit ITP, Chinese Academy of Sciences, Beijing, China for a visit where a part of the work is completed.

\end{document}